\documentclass[twocolumn,showpacs,preprintnumbers,amsmath,amssymb,prl]{revtex4}

\usepackage{epsfig}

\newcommand{\T}{\mathcal{T}}

\begin{document}

\title{Full counting statistics of spin transfer through ultrasmall quantum dots}
\author{T.~L.~Schmidt,$^{1}$ A.~Komnik,$^{1,2}$ and A.~O.~Gogolin$^3$ }

\affiliation{
\centerline{${}^1$Physikalisches Institut, Universit\"at Freiburg, D--79104 Freiburg, Germany} \\
\centerline{${}^2$Institut f\"ur Theoretische Physik, Universit\"at Heidelberg, D--69120 Heidelberg, Germany} \\
\centerline{${}^3$Department of Mathematics, 180 Queen's Gate, London SW7 2AZ, United Kingdom}}

\date{\today}

\begin{abstract}
We analyze the spin-resolved full counting statistics of electron
transfer through an ultrasmall quantum dot coupled to metallic
electrodes. Modelling the setup by the Anderson Hamiltonian, we
explicitly take into account the onsite Coulomb repulsion $U$. We
calculate the cumulant generating function for the probability to
transfer a certain number of electrons with a \emph{preselected}
spin orientation during a fixed time interval. With the cumulant
generating function at hand we are then able to calculate the spin
current correlations which are of outmost importance in the emerging
field of spintronics. We confirm the existing results for the charge
statistics and report the discovery of the new type of correlation
between the spin-up and -down polarized electrons flows, which has a
potential to become a powerful new instrument for the investigation
of the Kondo effect in nanostructures.

\end{abstract}


\maketitle


Modern microelectronics is one of the most successful technologies
ever conceived by humankind. However, in recent years a number of
limitations, which can slow down or even stop further progress began
to come to the fore. One possible way to overcome these difficulties
is to switch from charge current processing to spin current and spin
configuration processing. Their advantages are so enormous that
recently a completely new scientific field of spintronics has been
established \cite{awschalom02,molspintronics}.

In the conventional microelectronics the properties of the basic
circuitry elements are characterized by a number of different
quantities -- by the nonlinear current-voltage relations, by the
current noise spectra, by the current correlations of third order
(third cumulant) etc \cite{blanterbuettiker,blanter}. However, there
is one characteristic which (at least in the low frequency range)
contains information about correlations of \emph{all} orders. This
is the so-called full counting statistics (FCS), which answers all
above questions by providing the probability distribution $P(Q)$ to
transfer a certain amount of charge $Q$ during a waiting time
interval ${\cal T}$ \cite{levitov93,levitov96}.

While there is by now a vast amount of literature available on the
charge transfer statistics, its spin-resolved relative remains to
a larger part unknown with some notable exceptions
[\onlinecite{lorenzo04,kindermann05,schmidt07_new,martin07}]. We
would like to close this gap and analyze the combined statistics
of spin and charge transfer through ultrasmall quantum dots with
genuine repulsive interactions. The distinctive feature of these
devices are their extremely small lateral dimensions which allow
for only very few energy levels to take part in the transport
processes. Typical realizations are nanoscale heterostructures on
the semiconductor basis, or individual molecules coupled to
metallic electrodes, which are even smaller
\cite{kouwenhoven,goldhaber-gordon,weisvonklitzing,reichert}. Both
types of systems became available only during the last decades and
are promising candidates to become basic building blocks of future
nanoelectronics and spintronics circuitry.

The archetype system to describe such structures is the
single-impurity Anderson model \cite{anderson,hewson93}. It consists
of a local fermionic level (also called dot level) which is filled
or emptied by $d^\dag_\sigma,d_\sigma$ creation and annihilation
operators and $N$ different electronic continua modeling the
electrodes via fermionic fields $\psi_{\alpha\sigma}(x)$,
\begin{equation}\label{H0}
    H_0 =  \sum_{\sigma=\pm} (\Delta + \sigma h) d^\dag_\sigma d_\sigma +
    \sum_{\alpha = 1}^N \sum_{\sigma=\pm} H_0[\psi_{\alpha\sigma}]
    \, ,
\end{equation}
where $\Delta + \sigma h =: \delta_\sigma$ is responsible for the
energy of the dot in the magnetic field $h = g \mu_B B/2$, when it
is occupied by an electron with spin orientation $\sigma$. The dot
level energy $\Delta$ is an additional parameter which can in
practice be changed by varying the voltage on the background gate
electrode. The coupling of dot and leads is achieved by a local
tunneling contribution
\begin{equation}\label{HT}
    H_T = \sum_\alpha \sum_\sigma \gamma_\alpha \left[
    d^\dag_\sigma \psi_{\alpha\sigma}(x=0) + \mbox{h.c.} \right] \, ,
\end{equation}
with energy independent tunneling amplitudes $\gamma_\alpha$. In
addition to these terms one has to take into account the
electrostatic repulsion
\begin{equation}\label{HU}
    H_U = U (d^\dag_\uparrow
d_\uparrow - n_0) (d^\dag_\downarrow d_\downarrow - n_0) \, ,
\end{equation}
reflecting the energetic cost $U$ of double dot occupation with
respect to the symmetric value $n_0=1/2$. Due to this normalization,
the particle-hole symmetric case corresponds to $\Delta = 0$. The
full system Hamiltonian is the sum of all three contributions $H=
H_0 + H_T + H_U$.

The technology for the calculation of the FCS is by now far advanced
and allows for a number of different approaches. In the most
widespread one, the quantity of interest is the so-called cumulant
generating function (CGF) $\ln \chi(\lambda_\downarrow,
\lambda_\uparrow)=\ln \chi(\lambda_\sigma)$
\cite{levitov96,nazarov03,schoenhammer07}. Its successive
differentiation with respect to the counting fields $\lambda_\sigma$
yields the respective irreducible momenta $\langle \! \langle \delta
Q^n_\sigma \rangle \! \rangle$ for the probability distribution to
transfer $\delta Q_\sigma$ charges with spin orientation $\sigma$
through the system during the waiting time ${\cal T}$,
\begin{eqnarray}
    \langle \! \langle \delta Q^n_\sigma \rangle \! \rangle = (-i)^n
 \frac{\partial^n}{\partial \lambda^n_\sigma} \ln \,
 \chi(\lambda_\sigma) \big|_{\lambda=0}\, .
\end{eqnarray}
In analogy to the approaches taken in
[\onlinecite{levitov04,gogolin06}], the first step in the
calculation of the CGF is to endow the tunneling Hamiltonian
(\ref{HT}) with counting fields $\lambda_\sigma$. Outside of the
waiting time interval $0 < t < {\cal T}$ the counting fields are
zero. The tunneling Hamiltonian then transforms to


\begin{equation}\label{Tlambda}
 H_T \rightarrow H_T^\lambda = \sum_\alpha \sum_\sigma \gamma_\alpha
    \left[e^{i\lambda_{\alpha\sigma}/2} d^\dag_\sigma \psi_{\alpha\sigma} +
    \mbox{h.c.} \right] .
\end{equation}
In the noninteracting case ($U=0$, resonant level model
\cite{dejong,gogolin06}) this quantity can easily be calculated by
resummation of the perturbation series in $\gamma_{\alpha}$ or by
applying the Levitov-Lesovik formula \cite{levitov93}. Assuming
the measurement time $\T$ to be large such that switching effects
can be neglected, one finds (setting $e = \hbar = 1$)
\begin{eqnarray} \label{chi0}
    \ln \chi_0(\lambda_\sigma) &=& \T \sum_\sigma \int
    \frac{d\omega}{2\pi} \ln \Big\{
    1 + \sum_{\alpha\beta} T^{\alpha\beta}_\sigma(\omega)
    n_\alpha(1-n_\beta) \nonumber \\
&\times&
     \big[
    e^{i(\lambda_{\alpha\sigma}-\lambda_{\beta\sigma})} - 1\big]
    \Big\}\, ,
\end{eqnarray}
where $n_\beta(\omega)$ denotes the Fermi distribution in lead
$\beta$. The energy dependent transmission coefficients are given by
\begin{equation}\label{Talphabeta}
    T^{\alpha\beta}_\sigma(\omega) =
    \frac{4\Gamma_\alpha\Gamma_\beta}
    {(\omega-\delta_\sigma)^2 + \Gamma^2}\, ,
\end{equation}
where $\Gamma_\beta = \pi \nu \gamma_\beta^2$ (with the density of
states $\nu$ at the Fermi level) is the hybridization of the dot
with lead $\beta$ and $\Gamma = \sum_\beta \Gamma_\beta$.

While the non-interacting result (\ref{chi0}) was derived for an
arbitrary number $N$ of fermionic leads, we shall restrict
ourselves henceforth to the symmetric two level case, i.~e., we
assume $N=2$ and $\Gamma_L = \Gamma_R$. The chemical potentials of
the leads are assumed to be at $\mu_{L,R}$ where $V = \mu_L -
\mu_R$ denotes the applied voltage.

It is quite inefficient to calculate $\chi(\lambda_\sigma)$ using
an additional expansion in $U$ in the same way. As has been
realized in Ref.~[\onlinecite{gogolin06_2}] in a different
context, as long as one is only interested in small $U$, the CGF
can be calculated by a simple linked cluster like calculation.
Thereby the full $\lambda_\sigma$ dependence can be shifted onto
the unperturbed Keldysh Green's functions $D(\omega)$ of the dot
level. Using the notation of [\onlinecite{LLX}], these are given
by
\begin{eqnarray}\label{D}
D^{--}(\omega)&=& \left[ (\omega-\delta_\sigma) + \sum 2 i
\Gamma_\beta (n_\beta - 1/2)\right]/{\cal D}(\omega) \, , \nonumber \\
D^{-+}(\omega)&=& \left[ \sum 2 i \Gamma_\beta e^{
i\lambda_{\beta\sigma}} n_\beta
\right]/{\cal D}(\omega) \, , \nonumber \\
D^{+-}(\omega) &=& \left[ \sum 2 i \Gamma_\beta
e^{-i\lambda_{\beta\sigma}} (n_\beta - 1) \right]/{\cal D}(\omega)
\, , \nonumber \\
D^{++}(\omega) & = & - [ D^{--}(\omega) ]^* \, ,
\end{eqnarray}
where we defined
\begin{eqnarray}
{\cal D}(\omega) &=&
 (\omega-\delta_\sigma)^2 + \Gamma^2  \\
    &+& 4 \sum_{\alpha\beta} \Gamma_\alpha \Gamma_\beta
    n_\alpha(1-n_\beta) \left[ e^{i(\lambda_{\alpha\sigma} -
    \lambda_{\beta\sigma})} - 1\right] \, .\nonumber
\end{eqnarray}
Note that the presence of the counting field causes a violation of
Keldysh's sum rule. Using a linked cluster expansion, the exact CGF
$\ln \chi(\lambda_\sigma)$ can be expressed as a correction to the
noninteracting CGF (\ref{chi0}). The quantity we have to evaluate is
then
\begin{equation}
    \chi(\lambda_\sigma) = \chi_0(\lambda_\sigma) \big\langle T_C \exp\big[-i \int_C
    dt \ H_U(t)\big] \big\rangle \, .
\end{equation}
The expectation value is to be taken with respect to the
noninteracting ground state and therefore contains the
$\lambda$-dependent Green's functions (\ref{D}). It may appear that
since the lowest order expansion in $U$ only contains the dot
occupation numbers $n_\sigma$, one is not expecting any counting
field dependence to survive. However, this is no longer valid in the
case of explicitly (quite artificially) time-dependent
$\lambda_\sigma$. In the limit
\begin{eqnarray}                       \label{validity}
  \mbox{max} \{\Delta,h,V\}/\Gamma \ll 1 \, ,
\end{eqnarray}
the first order contribution is given by
\begin{eqnarray} \label{chi1_final}
 \ln \chi^{(1)}(\lambda_\sigma) = -\frac{U \T V}{\pi^2 \Gamma^3}
 \sum_\sigma \delta_\sigma \delta_{\bar{\sigma}} (e^{-i\lambda_\sigma} - 1)
 \, ,
\end{eqnarray}
where $\bar{\sigma} = - \sigma$. As it depends on $\delta_\sigma$,
this term only contributes for finite magnetic field ($h \neq 0$)
and/or broken electron-hole symmetry ($\Delta \neq 0$).

\begin{figure}[t]
    \centering
    \begin{tabular}{cc}
        \raisebox{10mm}{\begin{tabular}{c}
            \includegraphics[width=4cm]{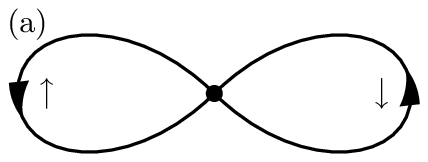} \\
            \includegraphics[width=4cm]{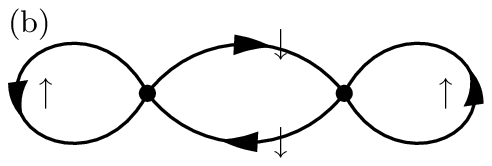}
        \end{tabular}} &
        \includegraphics[width=4cm]{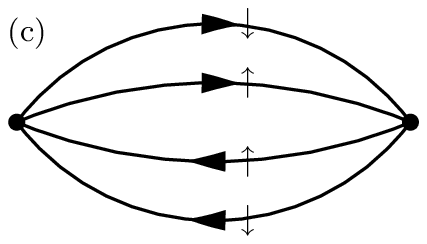}
    \end{tabular}
    \caption{One first order (a) and two second order (b,c) diagrams.}
    \label{FigU}
\end{figure}

The second order contribution is given by two different diagrams,
see Fig.~\ref{FigU}. One of these is again proportional to the
magnetic field and contains the average dot occupation numbers while
the other one is the double shell diagram. The calculation procedure
is rather lengthy but straightforward and results in the following
CGF expansion in the limit (\ref{validity}),
\begin{widetext}
\begin{eqnarray}\label{chi2_final}
    \ln \chi^{(2)}(\lambda_\sigma)
& = &
    \frac{\T V \chi_o^2}{2\pi \Gamma^2} \sum_\sigma \delta_{\bar{\sigma}}^2 ( e^{-i\lambda_\sigma} - 1) +
    \frac{\T V (\chi_e^2-1)}{2\pi \Gamma^2} \sum_\sigma \delta_\sigma^2 ( e^{-i\lambda_\sigma} - 1) \nonumber \\
& + &
    \frac{\T V^3 \chi_o^2}{24\pi \Gamma^2} \bigg\{
    4 (e^{-i\lambda_\uparrow-i\lambda_\downarrow} -1) +
    \sum_{\sigma} (e^{-i\lambda_\sigma} - 1) \bigg\} 
 + 
    \frac{\T V^3 (\chi_e^2 - 1)}{24\pi \Gamma^2}\sum_\sigma 
    (e^{-i\lambda_\sigma} - 1) \, , 
\end{eqnarray}
\end{widetext}
where we introduced the equilibrium even/odd susceptibilities
(correlations of $n_\uparrow$ with $n_\uparrow$ and $n_\downarrow$,
respectively), which are known to possess the following expansions
for small $U$ \cite{yamada1,yamada2,yamada3},
\begin{eqnarray}
    \chi_e = 1 + \left(3-\frac{\pi^2}{4} \right)
    \frac{U^2}{\pi^2\Gamma^2} \, \, \, , \, \, \,
    \chi_o = -\frac{U}{\pi\Gamma} \, .
\end{eqnarray}
Following the reasoning along the lines of
Ref.~[\onlinecite{gogolin06}], we may speculate that
(\ref{chi2_final}) is the exact result to all orders of $U$ and
$\Gamma$ as soon as one inserts the exact values for $\chi_{e,o}$
which have been obtained by, e.g., Bethe ansatz calculations
\cite{kawakamiokiji,vigmantsvelik}.
Now we are in a position to establish contact to known results and
to discuss new effects. Thus far, similar results have been obtained
only for the charge transport statistics of the same system for the
much more restrictive particle-hole symmetric parameter
constellation \cite{gogolin06}. The complete spin resolved
statistics for large transmission $\Gamma$ is given by the following
CGF,
\begin{widetext}
\begin{eqnarray}\label{chi_symm}
     \ln \chi(\lambda_\sigma)
& = &
    i G_0 \T V (\lambda_\uparrow + \lambda_\downarrow)
+
    \frac{\T V}{2\pi\Gamma^2} \sum_\sigma (\chi_e \delta_\sigma
    + \chi_o \delta_{\bar{\sigma}})^2(e^{-i\lambda_\sigma} - 1)
    \nonumber \\
& + &
    \frac{\chi_o^2 \T V^3}{6\pi\Gamma^2}
    (e^{-i\lambda_\uparrow-i\lambda_\downarrow}-1)
+
    \frac{(\chi_e^2 + \chi_o^2) \T V^3}{24\pi\Gamma^2}
    \sum_{\sigma} (e^{-i\lambda_\sigma}-1) \, ,
\end{eqnarray}
\end{widetext}
where $G_0 = 1/(2 \pi)$ is the conductance quantum per spin
orientation. In order to go over to the pure charge CGF one has to
set $\lambda_\uparrow = \lambda_\downarrow = \lambda$. In this case,
the result of Ref.~[\onlinecite{gogolin06}] is perfectly reproduced
for $h=\Delta=0$. Moreover, it had been speculated that while the
terms containing a single $\lambda$ correspond to single electron
tunneling events, the term with the doubled counting field is
brought about by a coherent electron pair tunneling in a singlet
state. Eq.~(\ref{chi_symm}) represents the proof of this conjecture
since the term giving rise to $2 \lambda$ part indeed stems from the
contribution originally containing the sum $\lambda_\uparrow +
\lambda_\downarrow$.

Yet another justification of the validity of (\ref{chi_symm}) for
arbitrary $U$ is brought about by comparing the above result to the
spin-resolved statistics of charge transfer through a Kondo impurity
in the unitary limit presented in [\onlinecite{gogolin06_2}].
Similar to the parameter mapping for the conventional current
statistics one identifies the two limits of small and large $U$
(Kondo regime) by
\begin{eqnarray}                    \label{constanttranslation}
\phi/T_K = \chi_o/\Gamma \quad\text{and}\quad \alpha/T_K =
\chi_e/\Gamma \, ,
\end{eqnarray}
where $T_K$ is the Kondo temperature and $\phi$ and $\alpha$ are
Fermi liquid parameters of the Kondo fixed point \cite{nozieres74}.
The similarity of these two results can be traced back to the
similarity of the corresponding Hamilton operators, which not only
both contain a resonant level part but also possess analogous
interaction terms.

Next, we would like to discuss the linear response (linear in $V$)
contribution. It can easily be verified that under the conditions
(\ref{validity}) one obtains the following CGF,
\begin{eqnarray}
    && \ln \chi(\lambda_\sigma)_{lin} = \\ \nonumber
   && G_0 \T V \sum_\sigma \ln \bigg\{ 1 + \frac{\Gamma^2}{(\chi_c \Delta +
    \sigma \chi_s h)^2 + \Gamma^2}
    (e^{i\lambda_\sigma} - 1) \bigg\} \, .
\end{eqnarray}
This result perfectly coincides with the conjecture of the
\emph{binomial theorem} formulated in [\onlinecite{gogolin06}]. It
predicts that the linear response charge transfer statistics of
\emph{any} interacting region coupled to noninteracting continua is
binomial and governed by the value of the transmission coefficient
at the Fermi edge. In fact, the road to the construction of the
spin-resolved CGF from Eq.~(26) of [\onlinecite{gogolin06}] (which
contains only the charge transfer generating function) is very
natural and intuitive: the logarithms with different signs in front
of the magnetic field term should contain counting fields for
different spin projections.

The spin current statistics can easily be recovered from the above
results after transition to the charge current and spin current
counting fields $\lambda$, $\mu$ via $\lambda_{\uparrow,
\downarrow} = \lambda \pm \mu$. One feature of (\ref{chi_symm}) is
the fact that the odd cumulants of spin currents are only non-zero
in finite field and for the particle-hole asymmetric case $\Delta
\neq 0$. This is the precise condition for the spin flow existence
in a conventional noninteracting resonant level system as well. In
the linear response regime the corresponding odd order cumulants
are then given by
\begin{eqnarray}
 \langle \! \langle (\delta Q_\uparrow - \delta Q_\downarrow)^{2n+1}
 \rangle \!
 \rangle =
 - \frac{2 \T V \Delta h}{\pi\Gamma^2} \, (\chi_e^2 - \chi_o^2) \, .
\end{eqnarray}
The even order cumulants are non-universal but $n$-independent as
well, so that the ratio of even/odd orders (it can be seen as a
generalization of the Fano factor) is given by
\begin{eqnarray}
 && \frac{\langle \! \langle (\delta Q_\uparrow - \delta Q_\downarrow)^{2n} \rangle \! \rangle}
 {\langle \! \langle (\delta Q_\uparrow - \delta Q_\downarrow)^{2n+1} \rangle \! \rangle} =
 \\ \nonumber &=&
 - \frac{\Delta^2 (\chi_e + \chi_o)^2 + h^2 (\chi_e - \chi_o)^2}{2  \Delta h (\chi_e^2
 - \chi_o^2)} \, .
\end{eqnarray}

Going beyond the linear response regime, we find that the most
fundamental feature emerging from (\ref{chi_symm}) is the
existence of the \emph{invariant cross-cumulant},
\begin{eqnarray}         \label{crosscumulant}
 \langle \! \langle \delta Q_\uparrow^n \delta Q_\downarrow^m \rangle \! \rangle &=&
 (-i)^{n+m} \, \frac{\partial^{n+m}}{\partial \lambda_\uparrow^n
 \partial \lambda_\downarrow^m} \, \ln  \chi(\lambda_\sigma)
 \nonumber \\
 &=&  (-1)^{n+m} \frac{\chi_o^2 \T V^3}{6\pi\Gamma^2}  \, ,
\end{eqnarray}
for $n,m \ge 1$. Not only is this quantity non-zero in interacting
systems only, it is also independent of magnetic field strength and
(up to the sign) of its orders $n,m$. It exists in the strong
coupling Kondo case as well and is found using
(\ref{constanttranslation}) for the parameter translation between
weak and strong coupling. Despite the formally identical
mathematical shapes, the origin of this phenomenon is completely
different for small $U$ and in the Kondo regime. While in the weak
coupling case it signifies the beginning of the spin singlet
formation, in the strong coupling limit it starts to appear as soon
as it becomes possible to break up (though virtually) the Kondo spin
singlet. The amplitude of these correlations grows as one approaches
the strong coupling fixed point. In principle, in addition to the
conventional linear conductance (which approaches the unitary limit
of almost perfect conductance) the cross-cumulant can also be
regarded as a measure of how deep in the Kondo regime the system in
question is being.

To conclude, we have analyzed the non-equilibrium spin resolved
FCS of the Anderson impurity model by calculating the CGF of the
probability distribution to transfer a fixed amount of charge with
preselected spin orientation during very long waiting time
interval. Our results perfectly agree with existing predictions
for the statistics of charge transfer. The CGF indeed supports the
interpretation that the electron transport in such a system is
mediated not only by single charge tunneling but by correlated
transfer of electron pairs in a singlet state. Moreover, the
emerging expressions confirm the previously conjectured statistics
in finite field beyond the particle-hole symmetric situation. In
the linear response regime, it turns out to be binomial and to
factorize in different spin channels. Finally, we have discovered
a new type of correlation between the spin-up and spin-down
currents: a cross--cumulant. It is universal and field
independent. In our view, it has a potential to become one of the
quantities to measure and control the `quality' of the Kondo
effect in nanostructures.

We would like to thank Hermann Grabert and Dario Bercioux for many
valuable discussions. This work was supported by DFG under grant
No. KO 2235/2 (Germany).

\bibliography{bibliography}

\end{document}